\documentclass[prl,twocolumn,showpacs,preprintnumbers,amsmath,amssymb,superscriptaddress]{revtex4}
\usepackage{graphicx}
\usepackage{dcolumn}
\usepackage{bm}
\usepackage{color}
\def\bvec#1{\mbox{\boldmath $#1$}}
\begin{document}


\title{\boldmath Analyzing power for the proton elastic scattering from
neutron-rich ${\rm ^6He}$ nucleus}

\author{T.~Uesaka}\email{uesaka@cns.s.u-tokyo.ac.jp}
\affiliation{Center for Nuclear Study, University of Tokyo, Tokyo 113-0033, Japan}
\author{S.~Sakaguchi}
\affiliation{Center for Nuclear Study, University of Tokyo, Tokyo 113-0033, Japan}
\author{Y.~Iseri}
\affiliation{Department of Physics, Chiba-Keizai College, Chiba 263-0021, Japan}
\author{K.~Amos} 
\affiliation{School of Physics, University of Melbourne, Melbourne, Australia}
\author{N.~Aoi}
\affiliation{RIKEN Nishina Center, Saitama 351-0198, Japan}
\author{Y.~Hashimoto}
\affiliation{Department of Physics, Tokyo Institute of Technology, Tokyo 152-8551, Japan}
\author{E.~Hiyama}
\affiliation{RIKEN Nishina Center, Saitama 351-0198, Japan}
\author{M.~Ichikawa}
\affiliation{Cyclotron \& Radioisotope Center, Tohoku University, Miyagi 980-8578, Japan}
\author{Y.~Ichikawa}
\affiliation{Department of Physics, University of Tokyo, Tokyo 113-0033, Japan}
\author{S.~Ishikawa}
\affiliation{Science Research Center, Hosei University, Tokyo 102-8160, Japan}
\author{K.~Itoh}
\affiliation{Department of Physics, Saitama University, Saitama 338-8570, Japan}
\author{M.~Itoh}
\affiliation{Cyclotron \& Radioisotope Center, Tohoku University, Miyagi 980-8578, Japan}
\author{H.~Iwasaki}
\affiliation{Department of Physics, University of Tokyo, Tokyo 113-0033, Japan}
\author{S.~Karataglidis}
\affiliation{Department of Physics and Electronics, Rhodes University, P.O. Box 94 Grahamstown 6140, South Africa}
\author{T.~Kawabata}
\affiliation{Center for Nuclear Study, University of Tokyo, Tokyo 113-0033, Japan}
\author{T.~Kawahara}
\affiliation{Department of Physics, Toho University, Chiba 274-8510, Japan}
\author{H.~Kuboki}
\affiliation{Department of Physics, University of Tokyo, Tokyo 113-0033, Japan}
\author{Y.~Maeda}
\affiliation{Faculty of Engineering, University of Miyazaki, Miyazaki 889-2192, Japan}
\author{R.~Matsuo}
\affiliation{Cyclotron \& Radioisotope Center, Tohoku University, Miyagi 980-8578, Japan}
\author{T.~Nakao}
\affiliation{Department of Physics, University of Tokyo, Tokyo 113-0033, Japan}
\author{H.~Okamura}
\affiliation{Research Center for Nuclear Physics, Osaka University, Osaka 567-0047, Japan}
\author{H.~Sakai}
\affiliation{Department of Physics, University of Tokyo, Tokyo 113-0033, Japan}
\author{Y.~Sasamoto}
\affiliation{Center for Nuclear Study, University of Tokyo, Tokyo 113-0033, Japan}
\author{M.~Sasano}
\affiliation{Department of Physics, University of Tokyo, Tokyo 113-0033, Japan}
\author{Y.~Satou}
\affiliation{Department of Physics, Tokyo Institute of Technology, Tokyo 152-8551, Japan}
\author{K.~Sekiguchi}
\affiliation{RIKEN Nishina Center, Saitama 351-0198, Japan}
\author{M.~Shinohara}
\affiliation{Department of Physics, Tokyo Institute of Technology, Tokyo 152-8551, Japan}
\author{K.~Suda}
\affiliation{Research Center for Nuclear Physics, Osaka University, Osaka 567-0047, Japan}
\author{D.~Suzuki}
\affiliation{Department of Physics, University of Tokyo, Tokyo 113-0033, Japan}
\author{Y.~Takahashi}
\affiliation{Department of Physics, University of Tokyo, Tokyo 113-0033, Japan}
\author{M.~Tanifuji}
\affiliation{Science Research Center, Hosei University, Tokyo 102-8160, Japan}
\author{A.~Tamii}
\affiliation{Research Center for Nuclear Physics, Osaka University, Osaka 567-0047, Japan}
\author{T.~Wakui}
\affiliation{Cyclotron \& Radioisotope Center, Tohoku University, Miyagi 980-8578, Japan}
\author{K.~Yako}
\affiliation{Department of Physics, University of Tokyo, Tokyo 113-0033, Japan}
\author{Y.~Yamamoto}
\affiliation{Tsuru University, Yamanashi 402-8555, Japan}
\author{M.~Yamaguchi}
\affiliation{RIKEN Nishina Center, Saitama 351-0198, Japan}

\date{\today}

\begin{abstract}
 Vector analyzing power for the proton-${\rm ^6He}$ elastic scattering
 at 71~MeV/nucleon has been measured for the first time, with a newly developed 
polarized proton solid target working at low magnetic field of 0.09~T. 
The results are found to be incompatible 
with a $t$-matrix folding model prediction. 
Comparisons of the data with $g$-matrix folding analyses clearly show
that the vector analyzing power is sensitive to the nuclear structure model used
in the reaction analysis. The $\alpha$-core distribution in ${\rm ^6He}$
is suggested to be a possible key to understand the nuclear structure sensitivity.
\end{abstract}

\pacs{24.70.+s, 25.60.-t, 29.25.Pj}
\maketitle

Spin observables in scattering experiments have been rich sources of our
understanding of nuclear structure, reaction, and interactions. 
One of the good examples is spin asymmetry in proton-proton and proton-nucleus ($p$-$A$) scatterings which 
is a direct manifestation of spin-orbit coupling in the system. The
first spin asymmetry measurements carried out by use
of a double scattering method~\cite{Oxley53,Chamberlain56} clearly
demonstrated that the spin-orbit coupling in nuclei is an order of
magnitude stronger than that due to the relativistic effect \cite{Fermi54}. At present
the spin-orbit coupling in $p$-$A$ scattering is quantitatively
established through numerous experiments using polarized proton beams for stable targets.

It is interesting to use spin asymmetry measurements to
study unstable nuclei. Nuclei locating near the neutron
drip line occasionally show distinctive structure such as halos or skins. 
The neutron rich ${\rm ^6He}$
nucleus is one of the typical nuclides with an extended neutron
distribution. 
Since the extended neutron distribution is prominent at the nuclear surface 
and the spin-orbit coupling is, in nature, a surface phenomenon, it is
stimulating to see how the extended neutron distributions affect the spin asymmetry,
i.e., vector analyzing power in proton elastic scattering. 

In this Letter, we report new results of vector analyzing power
for the $p$-${\rm ^6He}$ elastic scattering at 71~MeV/nucleon, measured with a newly
developed polarized proton target. The results are compared with 
microscopic folding model calculations.

Although cross sections in proton elastic scattering from ${\rm ^6He}$ have been extensively measured over a wide range of energies~\cite{Korsheninnikov97,Wolski99,Cortina-Gil97,Lagoyannis01,Egelhof02,Alkhazov02},
until recently there had been no measurement of vector analyzing power.
Since unstable nuclei are produced as secondary beams, we need a polarized proton target, practically in the solid state, for the spin-asymmetry studies.
In addition, the solid polarized proton target should work under a low magnetic
field of $B\sim$ 0.1~T for detection of
recoiled protons with magnetic rigidity as low as 0.3~Tm. The traditional
dynamical nuclear polarization technique~\cite{Goertz02}, demanding a magnetic field
higher than a few Tesla, can not be applied therefore. Although this difficulty might be 
overcome by applying a ``spin frozen'' operation, efforts to do so have not been 
successful so far. An alternative approach to overcome the problem is to develop a polarized target based on a new
principle which is independent of magnetic field strength.

We have succeeded in constructing a new solid polarized proton target working at a low magnetic field of about 
0.1~T~\cite{Wakui-PST05}. Here, protons in the target are
polarized by transferring electron polarization in photo-excited triplet
states of pentacene molecules via cross
polarization\cite{Henstra88}. The magnitude of the electron polarization
is 73\% and depends neither on the
magnetic field strength nor on the temperature of material. 
This makes it possible to operate the polarized target
under a low magnetic field of 0.1~T and a high temperature of 100~K. 

The first experiment with this target system was carried out in 2003 where spin asymmetry 
in the $p$-${\rm ^6He}$ elastic scattering was measured\cite{Hatano05}. The data presented interesting features which were completely incompatible with theoretical predictions. From phenomenological optical model analyses, it was implied that the $p$-${\rm ^6He}$ spin-orbit potential might extend to a larger radius compared with the $p$-${\rm ^6Li}$ case. In Ref.\cite{Crespo07}, Crespo and Moro claim that an extended neutron distribution cannot be responsible for the large spin-orbit radius. Thus connection between spin-orbit potential in the $p$-${\rm ^6He}$ scattering and the extended neutron distribution in ${\rm ^6He}$ is still unclear. However, accuracy of the previous data is insufficient 
for further detailed and quantitative analysis. This is mainly because the analyzing power data were obtained with an assumed value of target polarization\cite{Hatano05}.  

To obtain accurate analyzing power data with a reliable normalization, 
we have performed the $p$-${\rm ^6He}$ spin-asymmetry experiment with an upgraded target and detector systems.

\begin{figure}[htbp]
  \centering
  \resizebox{8.6cm}{!}{\includegraphics{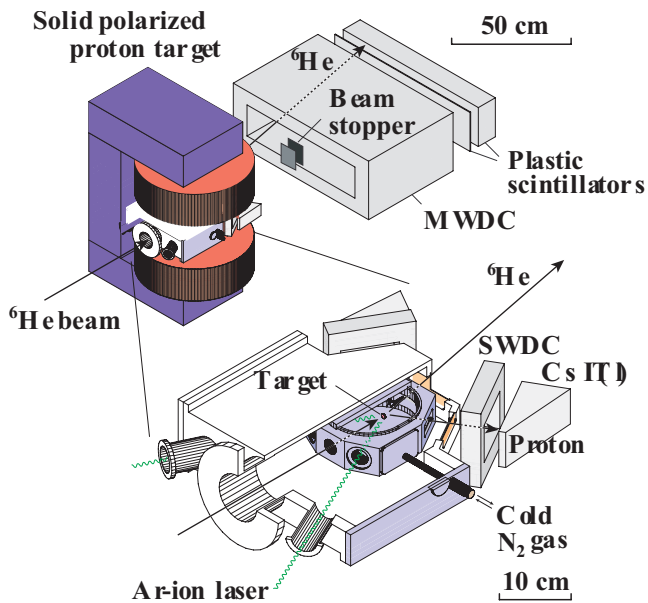}}
   \caption{(color online) Schematic layout of the experimental setup, including the polarized proton solid target. \label{fig:setup}}
\end{figure}

The experiment was performed at RIKEN Accelerator Research Facility. The
setup including the polarized target system, shown
in Fig.~\ref{fig:setup}, was placed downstream of the final focal plane of
RIPS~\cite{Kubo92}.  

A radioactive ${\rm ^6He}$ beam with an energy of 70.6$\pm 1.4$~MeV/nucleon was produced
via the projectile fragmentation reaction of a primary ${\rm ^{12}C}$
beam on a 1.39 g/cm$^2$ beryllium target. An energy and an average
intensity of the primary beam were 92~MeV/nucleon and 600~pnA, respectively.
The resulting intensity and purity of ${\rm ^6He}$ beam was 3.0$\times 10^{5}$~cps and 95\%, respectively.

After separation in RIPS, the ${\rm ^6He}$ beam bombarded the polarized target made of a crystal of naphthalene
with a small amount ($\sim$ 0.005~mol\%) of pentacene as a
dopant.  The dimensions of the target were 1~mm in thickness and 14~mm in diameter. 
The target was placed in a homogeneous magnetic field of 0.09~T produced by a
C-type magnet. The target chamber which was thermally isolated from the room-temperature environment and was
cooled down to 100~K by blowing a cold nitrogen gas. Laser light from
Ar-ion lasers irradiated the target to polarize electrons in pentacene molecules\cite{Wakui05}. 

The relative magnitude of the proton polarization was
monitored with a pulse NMR method during the measurement. 
The absolute value of the polarization was calibrated by comparing the NMR signal amplitude to the
asymmetry of the $p$-${\rm ^4He}$ scattering, measured with the same setup at 80~MeV/nucleon.
The analyzing power data for the $p$-${\rm ^4He}$ scattering in Ref.~\cite{Togawa87} were used in the calibration. The proton
polarization was found to be 20$\pm$4\% at maximum and 14$\pm$3\% on average. 
Statistical uncertainty in the $p$-${\rm ^4He}$ measurement dominates the uncertainties in the proton polarization.

Scattered ${\rm ^6He}$ particles were detected by a multi-wire drift
chamber (MWDC) and plastic scintillators placed about 1~m downstream of
the target. Pulse height information from the plastic scintillators is
used to identify the particle. The ${\rm ^6He}$ trajectory determined by
MWDC provides the scattering angle of the ${\rm
^6He}$ and the reaction position on the target. Two counter telescopes to detect
recoiled protons were placed left and right with respect to the beam axis. Each telescope consisted of a single-wire drift chamber (SWDC)
for a position measurement and a CsI(Tl) scintillator for
a total energy measurement. They covered an angular range of
$\theta_{\rm c.m.}=35^{\circ}$ -- 90$^{\circ}$ in the center of mass system.  
The background around the elastic scattering peak was reasonably small,
which enables us to obtain yields of interest reliably. 


\begin{figure}[htbp]
 \centering
 \resizebox{8cm}{!}{\includegraphics{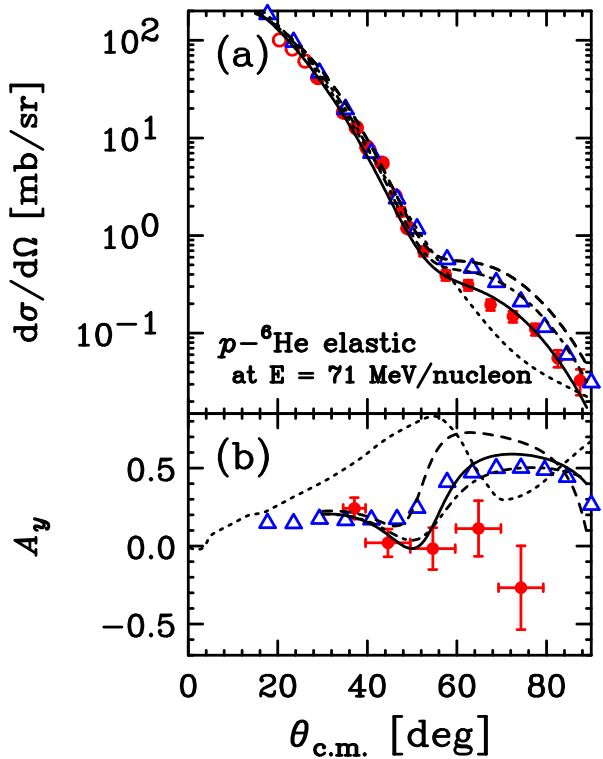}}
 \caption{(color online) Cross section and vector analyzing power for the $p$-${\rm ^6He}$ 
elastic scattering at 71~MeV/nucleon (filled circles), together with cross section data
 in Ref.~\cite{Korsheninnikov97} (open circles) and data for ${\rm ^6Li}$~\protect\cite{Henneck94} (triangles) targets. Dotted lines is a $t$-folding calculation in Ref.~\cite{Weppner00}. Results of 6BF calculations with harmonic oscillator (dashed), WS with (solid) and without halo (dot-dashed) single particle wave functions are shown. \label{fig:data}}
\end{figure}

The angular distributions of differential cross section
($d\sigma/d\Omega$) and vector analyzing power ($A_{y}$) are shown by filled circles in Fig.~\ref{fig:data}. 
Only statistical uncertainties are shown in the figure. Systematic uncertainty in $A_{y}$, mainly
due to uncertainty in absolute normalization of proton polarization,
is 19\% independent of scattering angles. 
The cross section data are obtained with systematic uncertainty of 9\% up to backward
angles of $\theta_{\rm c.m.} \le 87^{\circ}$.

In the top panel of the figure, the data
for the $p$-$\mathrm{^6Li}$ elastic scattering~\cite{Henneck94} (triangles) are shown for
comparison. As is indicated in Ref.\cite{Hatano05}, $d\sigma/d\Omega$ in the $p$-$\mathrm{^6He}$ scattering is almost identical to that in the $p$-$\mathrm{^6Li}$
scattering at $\theta_{\rm c.m.} \le 50^{\circ}$.
This indicates that matter distributions in
${\rm ^6He}$ and ${\rm ^6Li}$ are similar, which is consistent with the
recent results from GSI~\cite{Egelhof02}.
On the other hand, one can find small but apparent differences at backward angles $\theta_{\rm c.m.} > 50^{\circ}$,
which can be a manifestation of a halo structure in ${\rm ^{6}He}$\cite{Stepantsov02}.

In the lower panel of Fig.~\ref{fig:data}, we show the analyzing
power for the $p$-${\rm ^6He}$ elastic scattering at 71~MeV/nucleon,
together with the ${\rm ^6Li}$ data. In sharp contrast to the cross section, the $A_{y}$ data for the
$p$-$\mathrm{^6He}$ scattering are quite different from that for
$p$-$\mathrm{^6Li}$, especially at $\theta_{\rm c.m.} > 
50^{\circ}$. Dotted lines in Fig.~\ref{fig:data} represent prediction of the full $t$-folding
optical potential model by Weppner et al.~\cite{Weppner00} reported
before our measurement. This model calculates the $p$-$A$ scattering
amplitudes by folding free nucleon-nucleon scattering amplitudes ($t$-matrix) based on
the Nijmegen~I interaction with off-shell density matrices. The calculations predict large 
positive values of analyzing power in the region of measurement independent of the nuclear
structure model used. The predicted angular distribution is clearly inconsistent with the present $A_{y}$ data, while the calculation reproduces the
cross section data at forward angles reasonably. Those comparisons indicate that vector analyzing power can provide new information on the reaction mechanism and also on the nuclear structure, additional to that from the smaller scale effects in the elastic scattering cross section.

To obtain deeper understanding, we have compared the data with
two different $g$-matrix folding model calculations: one is a full six-body folding (6BF) calculation~\cite{Amos01,Stepantsov02} and the other is a
cluster-folding (CF) calculation which considers ${\rm ^6He}$ to have an explicit $\alpha$-core.

The $g$-matrix folding model, defined and used in Refs.~\cite{Amos01,Stepantsov02}, has
been successful in describing $p$-$A$ elastic and inelastic scatterings
for a wide range of nuclear masses and energies. In the model, the non-locality
of the $p$-$A$ interaction due to an exchange term is taken into account
in a fully microscopic way. Nuclear structure effects to the scattering 
are taken into account through single particle wave functions, one-body density matrix elements, and 
the $g$-matrix interaction. The $g$-matrix is obtained by solving the Bethe-Bruckner-Goldstone 
equations for Bonn-B potential. Use of the $g$-matrix is the largest difference from that in Ref.~\cite{Weppner00}. 
Three curves in Fig.~\ref{fig:data} represent results for different nuclear structure models: solid and dot-dashed curves are results with a single particle wave function for a Woods-Saxon (WS) potential with and without a halo component, respectively, while dashes curves are for a harmonic oscillator potential.
It is found that the predicted $A_{y}$ varies by as much as 0.2 depending on the nuclear structure model used in the analysis. Overall agreement to the $d\sigma/d\Omega$ and $A_{y}$ data can be obtained 
with WS wave functions. In particular $d\sigma/d\Omega$ at $\theta_{\rm c.m.}>50^{\circ}$
prefer a model with halo structure. 

What is the origin of this sensitivity to nuclear structure? Is it due to the {\it direct} valence neutron contribution or to the $\alpha$-core contribution, or both? Comparison with CF calculations is suited for clarifying the origin. Since the ${\rm ^6He}$ nucleus is known to have a
well-developed $\alpha$-$n$-$n$ structure, a folded interaction of
$p$-$\alpha$ and $p$-$n$ interactions with an $\alpha$-$n$-$n$ cluster
distribution should be a good approximation to the $p$-${\rm ^6He}$
interaction potential. The CF optical potential can be written as
$ U_{\rm CF} = \sum_{i=1,2}\int V_{pn_{i}}\rho_{n}(r_{i})d\bvec{r}_{i}
 +\int V_{p\alpha}\rho_{\alpha}(r_{\alpha})d\bvec{r}_{\alpha} $, 
where $V_{pX}$ includes both central and spin-orbit parts. 
In the actual calculation, a phenomenological optical potential which
reproduces the $p$-${\rm ^4He}$ elastic scattering
data at 72~MeV/nucleon~\cite{Burzynski89} is used as the $p$-$\alpha$
interaction. Complicated effects in the $p$-$\alpha$ interaction, 
such as non-locality due to exchange process, are considered to be 
simulated by the phenomenological optical potential, at least in part. 
The complex effective interaction (CEG)\cite{CEG} is adopted
as the $p$-$n$ interaction. Those interactions are folded with the $\alpha$-$n$-$n$ distributions determined by using Gaussian expansion method~\cite{Hiyama03}. Details of
the calculation will be reported elsewhere~\cite{Sakaguchi08}. 

\begin{figure}[htbp]
 \centering
 \resizebox{8cm}{!}{\includegraphics{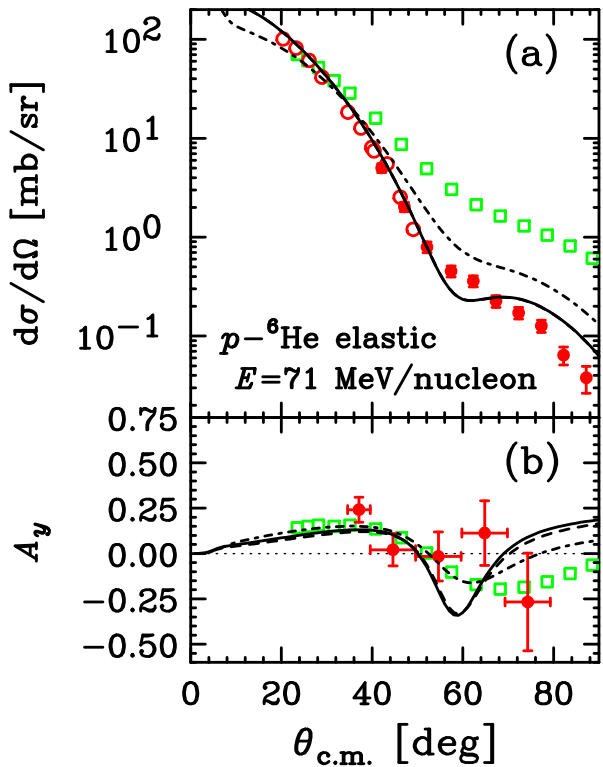}}
 \caption{(color online) Present data compared with the cluster-folding calculations. Solid, dashed, and dot-dashed lines represent calculations with full, $V_{pn;\ell s}=0$, and $V_{pn;\ell s}=V_{pn;{\rm central}}=0$ interactions, respectively. Data for the $p$-${\rm ^4He}$ scattering~\protect\cite{Burzynski89} (squares) are also shown. \label{fig:CF}}
\end{figure}

In Fig.~\ref{fig:CF}, results of the CF calculation (solid lines) are compared with the ${\rm ^6He}$ (circles) and ${\rm ^4He}$ (squares) ~\cite{Burzynski89} data. The ${\rm ^4He}$ data
are plotted at the angle where momentum transfer for $p$-${\rm ^6He}$ is the same as that for the corresponding $p$-${\rm ^4He}$ data. Although the angular distribution of $d\sigma/d\Omega$ for ${\rm ^6He}$ differs considerably from the more gradual one for ${\rm ^4He}$, the $A_{y}$ data are similar to each other.
The CF calculation reproduces both of $d\sigma/d\Omega$ and $A_{y}$ reasonably, in particular at $\theta_{\rm c.m.}\sim 35^{\circ}$--60$^{\circ}$. 

To separate the valence neutron and the $\alpha$-core contributions, calculations 
with $V_{pn;\ell s}=0$ (dashed lines) and with $V_{pn;\ell s}=V_{pn;{\rm central}}=0$ (dot-dashed lines)  have been made. The latter corresponds to extraction of a ``pure'' $\alpha$-core contribution.
As shown in Fig.~\ref{fig:CF}, the $p$-$n$ central interaction causes a sizable effect on $d\sigma/d\Omega$. 
Due to the $\alpha$-core motion in ${\rm ^6He}$, matter distribution of the core part is wider than that of a bare ${\rm ^4He}$ nucleus, while it is naturally narrower than that of ${\rm ^6He}$ as a whole. Reflecting this, $d\sigma/d\Omega$ for the $\alpha$-core contribution appears between ${\rm ^4He}$ and ${\rm ^6He}$ data. 
It is also found that the spin-orbit interaction $V_{pn;\ell s}$ gives negligible effects on
$d\sigma/d\Omega$ and $A_{y}$, which is consistent with predictions in Ref.~\cite{Crespo07}.
 Thus, from comparisons with CF, it is concluded that nuclear structure sensitivity of $A_{y}$
does not originate from the direct valence neutron contribution, but from the $\alpha$-core contribution.
The latter, which is affected by recoil of the valence neutrons, seems to be a possible key to understand the behavior of $A_{y}$. 

In summary, we have performed an experiment to measure $A_{y}$ and $d\sigma/d\Omega$
in scattering of a neutron-rich ${\rm ^6He}$ nucleus from protons, by using the newly-developed polarized proton solid target. The $A_{y}$ data are obtained with a proton polarization determined by asymmetry in the $p$-${\rm ^4He}$ scattering. 
We conclude, from comparisons with the microscopic folding model calculations, that (1) overall agreement between the present data and 6BF calculation is found for WS wave function. In particular, the $d\sigma/d\Omega$ data at backward angles favor existence of halo structure in ${\rm ^6He}$; (2) the data are reproduced by CF calculations reasonably well. The CF calculations show that direct contribution from the valence neutrons to analyzing power is negligibly small. Thus nuclear structure effects on $A_{y}$ may originate from the spatial distribution of the $\alpha$-core in ${\rm ^6He}$,
 which is closely connected to the valence neutron distribution. This can be a possible key to understand 
spin-orbit coupling in a neutron-rich ${\rm ^6He}$ nucleus.

The present work has demonstrated that the technique to polarize protons
in a low magnetic field can open a new possibility to explore physics of 
unstable nuclei. Experiments with the target would provide fruitful results in 
future radioactive nuclear beam facilities.

The authors thank the RIKEN and CNS staffs for operation of accelerators
during the measurement. One of the authors (S.S.) expresses his gratitude for financial support by a Grant-in-Aid for Japan Society for the Promotion of Science (JSPS) Fellows (No. 18-11398).
This work was supported by the Grant-in-Aid No. 17684005 of the Ministry
of Education, Culture, Sports, Science, and Technology of Japan.


\begin{thebibliography}{10}

\bibitem{Oxley53}
C.~Oxley {\it et~al.}, Phys. Rev. {\bf 91},  419  (1953).

\bibitem{Chamberlain56}
O.~Chamberlain {\it et~al.}, Phys. Rev. {\bf 102},  1659  (1956).

\bibitem{Fermi54}
E.~Fermi, Nuovo Cimento {\bf 10},  407  (1954).

\bibitem{Korsheninnikov97}
A.~Korsheninnikov {\it et~al.}, Nucl. Phys. A {\bf 617},  45  (1997).

\bibitem{Wolski99}
R.~Wolski {\it et~al.}, Phys. Lett. B {\bf 467},  8  (1999).

\bibitem{Cortina-Gil97}
M.~D.~Cortina-Gil {\it et~al.}, Phys. Lett. B {\bf 401},  9  (1997).

\bibitem{Lagoyannis01}
A.~Lagoyannis {\it et~al.}, Phys. Lett. B {\bf 518},  27  (2001).

\bibitem{Egelhof02}
P.~Egelhof {\it et~al.}, Eur. Phys. Jour. A {\bf 15},  27  (2002).

\bibitem{Alkhazov02}
G.~D.~Alkhazov {\it et~al.}, Nucl. Phys. A {\bf 712},  269  (2002).

\bibitem{Goertz02}
S.~Goertz, W. Meyer, and G. Reicherz, Prog. Part. Nucl. Phys. {\bf 49},  403  (2002), 
and references therein.

\bibitem{Wakui-PST05}
T. Wakui,  in {\em Proc. XIth Int. Workshop on Polarized Ion Source and
  Polarized Gas Targets 2005}, edited by T. Uesaka, H. Sakai, A. Yoshimi, and
  K. Asahi (World Scientific, Singapore, 2007), p.\ 49.

\bibitem{Henstra88}
A.~Henstra, P.~Dirksen, and W.~T.~Wenckebach, Phys. Lett. A {\bf 134},  134
   (1988).

\bibitem{Kubo92}
T.~Kubo {\it et~al.}, Nucl. Instr. Meth. B {\bf 70},  309  (1992).


\bibitem{Wakui05}
T.~Wakui {\it et~al.}, Nucl. Instr. Meth. A {\bf 550},  521  (2005).

\bibitem{Hatano05}
M.~Hatano {\it et~al.}, Eur. Phys. Jour. A {\bf 25},  255  (2005).

\bibitem{Crespo07}
R.~Crespo and A.~M.~Moro, Phys. Rev. C {\bf 76},  054607  (2007).

\bibitem{Togawa87}
H.~Togawa and H. Sakaguchi, RCNP Annual Report  1  (1987).

\bibitem{Uesaka04}
T.~Uesaka {\it et~al.}, Nucl. Instr. and Meth. A {\bf 526},  186  (2004).

\bibitem{Henneck94}
R.~Henneck {\it et~al.}, Nucl. Phys. A {\bf 571},  541  (1994).

\bibitem{Stepantsov02}
S.~V.~Stepantsov {\it et~al.}, Phys. Letts. {\bf B542}, 35 (2002).


\bibitem{Weppner00}
S.~P.~Weppner, O.~Garcia, and Ch.~Elster, Phys. Rev. C {\bf 61},  044601
  (2000).

\bibitem{Amos01}
K.~Amos {\it et~al.}, Adv. Nucl. Phys. {\bf 25},  275  (2001).

\bibitem{Burzynski89}
S.~Burzynski {\it et~al.}, Phys. Rev. C {\bf 39},  56  (1989).

\bibitem{CEG}
N.~Yamaguchi, S.~Nagata, and T.~Matsuda, Prog. Theor. Phys. {\bf
  70},  459  (1983).

\bibitem{Hiyama03}
E.~Hiyama, Y.~Kino, and M.~Kamimura, Prog. Part. Nucl. Phys.
  {\bf 51},  223  (2003).

\bibitem{Sakaguchi08}
S. Sakaguchi, Y.~Iseri {\it et~al.}, in preparation.


\end{thebibliography}

\end{document}